# Orthology: definitions, inference, and impact on species phylogeny inference

Rosa Fernández[1], Toni Gabaldón[1,2,3,*], Christophe Dessimoz[4,5,6,7,8,*]

[1]Bioinformatics and Genomics Unit, Centre for Genomic Regulation (CRG), The Barcelona Institute of Science and Technology, Dr. Aiguader 88, Barcelona 08003, Spain
[2]Universitat Pompeu Fabra (UPF). 08003 Barcelona, Spain
[3]ICREA, Pg. Lluís Companys 23, 08010 Barcelona, Spain
[4]Department of Computational Biology, University of Lausanne, Lausanne, Switzerland
[5]Center for Integrative Genomics, University of Lausanne, Lausanne, Switzerland
[6]Swiss Institute of Bioinformatics, Genopode Building, Lausanne, Switzerland
[7]Centre for Life's Origins and Evolution, Department of Genetics, Evolution & Environment, University College London, London, UK
[8]Department of Computer Science, University College London, London, UK

*Correspondence: Toni Gabaldón (toni.gabaldon@crg.eu) or Christophe Dessimoz (c.dessimoz@ucl.ac.uk)

**Abstract:** Orthology is a central concept in evolutionary and comparative genomics, used to relate corresponding genes in different species. In particular, orthologs are needed to infer species trees. In this chapter, we introduce the fundamental concepts of orthology relationships and orthologous groups, including some non-trivial (and thus commonly misunderstood) implications. Next, we review some of the main methods and resources used to identify orthologs. The final part of the chapter discusses the impact of orthology methods on species phylogeny inference, drawing lessons from several recent comparative studies.

## Introduction

All life on earth shares a common origin. An evidence for this is, for instance, the existence of "universal" genes shared by all living beings. Indeed, we can find genes that are so similar within or between species that we can infer to be evolutionarily related and share ancestry—i.e. *homologous*—beyond reasonable doubt. Identifying homologous genes is of great interest, because it is the first step toward identifying what is conserved, and what has changed during evolution. In addition, because experimental characterisation of genes remains labour intensive, assessing evolutionary relationships provides a way to interpolate or extrapolate gene attributes among different species, such as the structure and function of the proteins they encode (Eisen, 1998)—a goal for which understanding orthology relationships is key, as discussed below.

One key refinement is to try to distinguish more precisely *how* homologous genes are related, giving rise to different homology subtypes. Homologs arising through speciation are called *orthologs* (Fitch, 1970); those arising through duplication are called *paralogs* (Fitch, 1970); those arising through whole genome duplication (also referred to as *homopolyploidization* or *autopolyploidization* in plants) as *ohnologs* (Leveugle et al., 2003); those through hybridization followed by genome doubling (*allopolyploidization*) are referred to as *homoeologs* (Glover et al., 2016; Huskins, 1931); those through lateral gene transfer as *xenologs* (Gray and Fitch, 1983).

Here, we focus mostly in orthologs, which are of particular importance in phylogenomics as they provide the basis to infer species phylogenies. In the first part, we review more precisely how orthology is defined and inferred. We start with orthology between two species, and then consider orthology in multispecies contexts. In the second part, we discuss the impact of orthology on phylogenetic inference.

## Definitions, implications, complications

The term "ortholog" was coined by Walter Fitch nearly 50 years ago (Fitch, 1970):

> *"It is not sufficient, for example, when reconstructing a phylogeny from amino acid sequences that the proteins be homologous. […] there should be two subclasses of homology. Where the homology is the result of gene duplication so that both copies have descended side by side during the history of an organism (for example, α and β hemoglobin) the genes should be called paralogous (para = in parallel). Where the homology is the result of speciation so that the history of the gene reflects the history of the species (for example α hemoglobin in man and mouse) the genes should be*

...

> *called orthologous (ortho = exact). Phylogenies require orthologous, not paralogous, genes."*

The definition was visionary, eloquent, and seemingly simple to grasp. Yet there are several implications and complications that have lead to frequent misunderstanding and inconsistencies in the literature. We consider these in turn.

Fitch defines orthology and paralogy as relationships between two genes, depending on the type of initial evolutionary event that gave rise to the pair. This implies that subsequent events, e.g. duplications of one and/or the other gene have no bearing on the type of relationship. Such duplications can however mean that a gene can have more than one orthologous counterpart in another species. In other words, orthology can be not only a one-to-one relationship, but also a one-to-many, many-to-one or many-to-many relationship.

Furthermore, note that the definitions are indifferent to the position of the genes on the genome. Consider *e.g.* a mammal gene retained in the human lineage and duplicated in the rodent lineage. Consider furthermore that one mouse copy has remained in its ancestral locus and the other one has moved elsewhere in the genome. Both rodent paralogous copies are orthologous to the human gene. To specify a conserved locus, the concept of "positional ortholog" has been proposed (Dewey, 2011).

In Fitch's examples, the paralogs ("α- and β-hemoglobin") belong to the same organism while the orthologs ("α-hemoglobin in man and mouse") belong to different species. Is paralogy still meaningful when the two genes are found in two different species? The answer is a resolute "yes". For instance, α-hemoglobin in mouse and β-hemoglobin in human are paralogs because they resulted from a duplication in a common ancestor of the two species.

A more tricky question is the converse: is orthology still meaningful if the two genes belong to the same species? To answer this, we need to consider the possibility that two genes resulting from a speciation event end up inside the same organism. This is unusual, but could happen through lateral gene transfer or hybridisation. However, in such cases, a different terminology is customarily used—xenologs or homoeologs respectively. Calling such genes "orthologs" would be consistent with Fitch's definition, but would be at odd with common usage in the literature.

## From pairwise to groupwise orthology

Moving on beyond two genes, let us consider how orthology and paralogy apply to more than two species at a time. This generalisation is not a straightforward one because orthology and paralogy relationships are not transitive. That is, if gene A is orthologous to B, and B is orthologous to C, one cannot conclude that A and B are orthologous to each other. For instance, mouse has two insulin genes *Ins1* and *Ins2*, which duplicated within the rodent lineage *(Shiao et al., 2008)*. Human has one copy, *INS*. Therefore, *Ins1* is orthologous to *INS*, *INS* is orthologous to *Ins2*, but *Ins1* is not orthologous but paralogous to *Ins2*. The same is true for paralogous relationships.

The original Fitch definition is that of a pairwise relationship between two genes. However, the number of pairwise relationships grows quadratically with the number of genes and species considered. Moreover, as we have seen, there is no straightforward extrapolation of pairwise relationships among groups of genes or across species. This, together with the difficulty of expressing and interpreting pairwise relationships when referring to groups of genes in several species, prompted the concept of "orthologous groups".

Two main kinds of orthologous groups have been proposed (Fig. 1). One kind—which we refer to as "strict" orthologous groups—denotes sets of genes for which every two members are orthologous. This is the case for sets of one-to-one orthologs, as one-to-one orthology is transitive. More generally, it is also possible for such orthologous groups to span over duplication events as long as the resulting groups do not include any pair of paralogs. For instance, in the insulin example from above, a group containing *INS* and *Ins1* would fulfill this definition (Fig. 1).

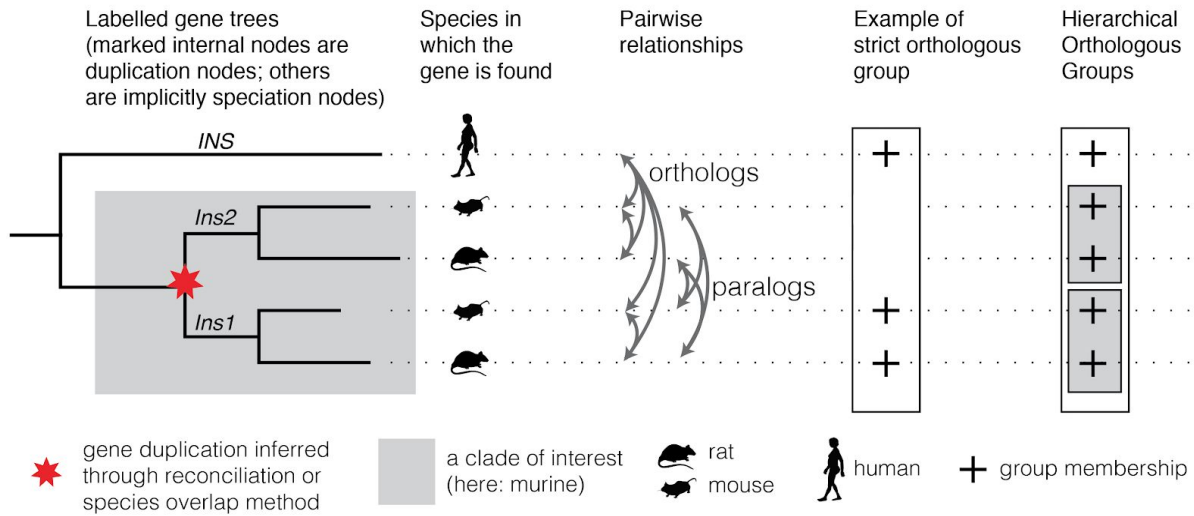

**Figure 1:** Conceptual overview of key concepts in this chapter. Gene tree with duplications and speciation nodes identified by reconciliation or species overlap; pairwise orthology and paralogy relationships; strict and hierarchical orthologous groups.

The other main kind of orthologous groups, called "hierarchical orthologous group" to avoid ambiguity, aims to identify sets of genes that have descended from a common ancestral gene in a given ancestral species. In the insulin example, since human *INS,* mouse *Ins1,* and mouse *Ins2* all descend from a common ancestral gene in the last common ancestor of all mammals, they are in a common hierarchical orthologous group at that level. By contrast, since *Ins1* and *Ins2* duplicated prior to the last murine common ancestor, these two genes are in different groups at the murine level (Fig 1). Thus, we can see that hierarchical orthologous groups are defined with respect to specific clades. Furthermore, we can see their *hierarchical* nature in that groups defined with respect to deeper clades subsume multiple groups defined on their descendants—as the insulin example also illustrates. This definition of orthologous groups is similar to the older concept of "subfamily" used to describe a subset of members of a gene family that share a common ancestry (i.e. form a clade in the gene family tree).

One special type of hierarchical groups is worth mentioning. When dealing with two species only, the point of reference is implicitly meant to be their last common ancestor. In this case, hierarchical orthologous groups coincide with "ortholog clusters" as defined by the method *Inparanoid* (Remm et al., 2001).

For a more in-depth review of the different types of orthologous groups, we refer readers to Boeckmann et al. (2011).

## Reconciled gene trees

As an alternative to orthologous groups, it is also possible to capture orthologous relationships using rooted gene trees which have their internal nodes labelled as speciation or duplication nodes (or possibly even more types of nodes). Such trees are commonly referred to as "labelled" or "reconciled" gene trees. All orthology and paralogy relationships among pairs of extant genes (i.e. pairs of leaves) in such trees can be deduced from the label associated with their last common ancestor: if the last common ancestor is a speciation node, the two genes are orthologs; if it is a duplication, they are paralogs (Fig. 1). Methods to infer the duplication and speciation labels are reviewed below.

Likewise, hierarchical orthologous groups can be obtained from the clades rooted in the speciation nodes corresponding to the taxonomic range of interest.

Therefore, labelled gene trees capture all orthology and orthologous group information. In addition, gene trees convey the order of gene duplications and quantify the amount of sequence divergence in the branch lengths.

Now that we have defined orthology, paralogy, orthologous groups, and reconciled trees, we turn to methods to infer these various types of relationships.

## State-of-the-art methods and resources

In this section, we provide an overview of methods and databases for orthology inference (Table 1). Methods are commonly divided into two main groups: tree-based and graph-based methods. Tree-based methods, as the name indicates, explicitly infer gene trees at some stage of their algorithms. By contrast, graph-based methods avoid inferring trees and instead compare sequences in a pairwise fashion, and build a graph with genes as vertices and some measure of sequence similarity as edges. We detail the two types of approach in turn, and refer to some of the popular associated algorithms and databases.

## Tree-based approaches

As we already discussed in the definition section, tree-based orthology inference methods reconstruct a gene tree for a group of homologous sequences to then infer the type of evolutionary event represented by each internal node of the tree. To infer events at internal nodes, the conventional approach is to perform "gene tree/species tree reconciliation". This can be done in a parsimony framework, e.g. Forester (Zmasek and Eddy, 2001) or Notung (Chen et al., 2000), or in a likelihood framework, e.g. GSR (Akerborg et al., 2009) or Phyldog (Boussau et al., 2013). Alternatively, the labelling of internal nodes can be determined by the method of species overlap, which labels as duplication node any internal node which has the same species represented in more than one of its child subtrees (Huerta-Cepas et al., 2007; van der Heijden et al., 2007). Thus the species overlap approach does not require or assume any species tree. Or, to be more precise, it considers a fully-unresolved species tree. Hence, it relies on the two copies that result from each gene duplication to be retained in at least one species, which is often the case in practice. This algorithm is considerably more robust to topological diversity in the gene trees—in contrast to gene/species tree reconciliation methods, which tend to introduce duplication events to explain any departure from the canonical species tree.

Several resources provide reconciled gene trees. PhylomeDB (Huerta-Cepas et al., 2014) and MetaPhOrs (Pryszcz et al., 2011) use the species overlap approach. For each reference species in the database, PhylomeDB infers a gene tree starting from each protein (each "seed"), and refers to the resulting set of trees as the *phylome* of that species. The species overlap method is also used and is available as part of the ETE software library (Huerta-Cepas et al., 2016a). Species overlap is extended to unrooted gene trees with UPhO (Ballesteros and Hormiga, 2016). PANTHER trees (Mi et al., 2017) infers reconciliation for all PANTHER families using the GIGA algorithm (Thomas, 2010), a gene/species reconciliation method. Likewise EnsemblCompara infers reconciled gene trees relating all Ensembl genomes using the TreeBeST algorithm (Vilella et al., 2008).

## Graph-based approaches

Graph-based approaches are based on comparisons between pairs of genes within and between species. They are all based on the observation that, for pairs of genes between two species, orthologs tend to be the pairs of sequences that have diverged the least. This is because until the speciation event that relates the two species, the orthologs were the same genes, while paralogs are the result of earlier duplications, and thus have more time to diverge.

This insight gave rise to the first large-scale orthology prediction approach, the basic "bidirectional best hit" (BBH) approach (Overbeek et al., 1999), which considers the pairs with mutually highest alignment scores, or its phylogenetic distance-based counterpart entitled "reciprocal shortest distance" (RSD) (Wall et al., 2003).

However, BBH and RSD do not deal well with many-to-many orthology relationships, resulting in missing pairs (Dalquen and Dessimoz, 2013). To address this, the Inparanoid algorithm provided a way to identify many-to-many orthology relationships (Remm et al., 2001). Furthermore, BBH and RSD can fail in case of differential gene loss—a situation where the corresponding ortholog is simply missing in both species, resulting in paralogs being wrongly identified as orthologs. The OMA algorithm introduced the use of third-party species, which might have retained both copies, which could thus act as "witnesses of non-orthology" (Dessimoz et al., 2006).

The other limitation of BBH and RSD is that they do not obviously generalise to groupwise orthology. The COGs database pioneered the use of "triangles" of pairwise orthologs (complemented by manual curation) to build multi-species orthologous groups (Tatusov et al., 1997). OrthoMCL used Markov clustering instead (Li et al., 2003). One issue with OrthoMCL is however that the granularity of the resulting groups depends on the choice of parameter ("inflation parameter"), which makes it harder to interpret the results.

The main graph-based resources include EggNOG (Huerta-Cepas et al., 2016b), HaMStR (Ebersberger et al., 2009), Inparanoid/HieranoiDB (Kaduk et al., 2017; Sonnhammer and Östlund, 2015), OMA (Altenhoff et al., 2018a), OrthoDB (Zdobnov et al., 2017), OrthoFinder (Emms and Kelly, 2015), and OrthoInspector (Nevers et al., 2019).

**Table 1:** Orthology methods mentioned in this chapter. For more methods, consult the *Quest for Orthologs* consortium website at https://questfororthologs.org/orthology_databases.

| Method | Type | Comments | Reference |
|---|---|---|---|
| BUSCO | Graph | Based on precomputed "universal single-copy" genes (defined for a number of standard clades), and thus inherently limited to these. Originally developed to assess genome completeness. | (Waterhouse et al., 2017) |
| COG/KOG | Graph | One of the first methods, still widely used for prokaryotic data. Includes a manual curation step. | (Tatusov et al., 2003) |
| EggNOG | Hybrid | Originally developed as extension of COG/KOG. Recent versions also include tree-based refinements. | (Huerta-Cepas et al., 2016b) |
| ETE 3.0 | Tree | General purpose tree analysis and visualisation package for Python, with species overlap function. | (Huerta-Cepas et al., 2016a) |
| Forester | Tree | General purpose tree analysis and visualisation software, including reconciliation function. | (Zmasek and Eddy, 2001) |
| GIGA | Tree | Gene/species tree reconciliation algorithm used in the PANTHER database. Also includes a heuristic for lateral gene transfer detection. | (Thomas, 2010) |
| GSR | Tree | Probabilistic gene/species tree reconciliation method | (Akerborg et al., 2009) |
| HaMSTR | Graph | The method uses a reference species to define one Hidden Markov Model per orthologous group, followed by reciprocal best hit within a family | (Ebersberger et al., 2009) |
| Hieranoid | Graph | Successor of Inparanoid to infer hierarchical orthologous groups from multiple species | (Kaduk et al., 2017) |
| Inparanoid | Graph | Infers orthologous groups independently for each pair of species. | (Sonnhammer and Östlund, 2015) |
| MetaPhOrs | Hybrid | Meta-method integrating predictions from multiple sources. | (Pryszcz et al., 2011) |
| Notung | Tree | Gene/species tree reconciliation software, with optional support for lateral gene transfer inference. | (Chen et al., 2000) |
| OMA | Graph | Infers both types of groups reviewed in this chapter: strict groups (suitable as markers for species tree inference) and hierarchical orthologous groups. | (Altenhoff et al., 2018a) |
| OrthoDB | Graph | Infers hierarchical orthologous groups. Used to infer the single-copy universal gene models of BUSCO. | (Zdobnov et al., 2017) |
| OrthoFinder | Graph | Infers hierarchical orthologous group with respect to the deepest speciation level only (the last common | (Emms and Kelly, 2015) |

| | | ancestor) | |
|---|---|---|---|
| OrthoInspector | Graph | Provides phylogenetic profiles as well. | (Nevers et al., 2019) |
| OrthoMCL | Graph | Groups inferred by OrthoMCL do not have a straightforward interpretation (they are neither strict nor hierarchical). Often used in combination with other methods and/or criteria. | (Li et al., 2003) |
| Phyldog | Tree | Gene/Species tree reconciliation in a maximum likelihood framework. Can be used to infer the species tree. | (Boussau et al., 2013) |
| PhylomeDB | Tree | Based on species overlap method. | (Huerta-Cepas et al., 2014) |
| UPhO | Tree | Species overlap method considering multiple gene tree rootings. | (Ballesteros and Hormiga, 2016) |

## Impact on phylogenetic inference: resolving the Tree of Life

Resolving the Tree of Life has been one of the prevailing questions in evolutionary biology at all systematic levels since the origin of phylogenetics. From bacteria to eukaryotes, from archaea to metazoa, great scientific efforts have been devoted towards understanding the evolutionary relationships between organisms.

The first sources of phylogenetic information to infer species trees were morphological characters. These characters were first classified as homologous or not based on taxonomic comparisons, then into ancestral or derived; finally phylogenetic interrelationships were inferred based on a parsimony criterium (Fitch, 1971). With the advent of molecular biology techniques, scientific efforts shifted largely to the use of molecular markers, which were aligned, concatenated (if several markers were used) and used to reconstruct a phylogeny. This approach would only provide sensible results if aligned sequences are orthologous to each other, as orthologs define speciation nodes, which constitute the only type of nodes that are expected in a species tree. If some of the sequences included have paralogous relationships, then some of the reconstructed nodes will indeed represent duplications and the resulting topology will be faulty with respect to the aimed species tree.

Initially, the experimental design in molecular phylogenetics included the identification of highly conserved regions in the organismal lineage of interest, that were amplified with specific probes by means of a polymerase chain reaction (PCR). As the same marker gene —i.e. the orthologous gene—was specifically sequenced from each of the species of interest, there was no need to search for orthologs. However, problems such as cross-amplification of paralogs, non-specific amplifications in the absence of the ortholog, or hidden paralogy issues, were common problems that could complicate the process of species tree reconstruction and have their root in the failure of obtaining a fully orthologous sequence dataset. With the advent of high-throughput sequencing and the availability of complete (or nearly complete) genomes and transcriptomes, one can in principle choose among virtually any marker gene. In these cases, there is a need of inferring orthologous genes from the source genomic datasets, and doing so correctly is pivotal for accurately reconstructing a species tree. As we will see below, despite the availability of automated methods, problems are likely to be encountered.

The past decade has seen an explosion of genome and transcriptome sequences from non-model organisms. More often than not, phylogenomic datasets include transcriptomes and low-coverage genomes that are incomplete, and contain errors and unresolved isoforms. These characteristics can severely violate the assumptions underlying some orthology inference methods. As a result, different orthology methods can result in very different phylogenetic inferences. Despite this fact, the effect of orthology inference is not commonly considered in typical phylogenomic analyses aimed at reconstructing species trees. Instead, methodological discussions have largely focused on the effect of phylogenetic reconstruction parameters such as the chosen models of substitution applied to the datasets, or on the effect of confounding factors, including missing data, compositional heterogeneity, or incomplete taxon sampling, among others. This is perhaps best epitomized by the intense debate around the position of ctenophores (Borowiec et al., 2015; Dunn et al., 2008; Hejnol et al., 2009; Moroz et al., 2014; Shen et al., 2017; Whelan et al., 2015) or sponges (Philippe et al., 2011, 2009; Pick et al., 2010; Pisani et al., 2015; Simion et al., 2017) as the earliest-branching phylum in the Animal Tree of Life.

Orthology benchmarking requires curated information about the underlying gene and species trees (e.g., the *Quest for Orthologs* benchmark service (Altenhoff et al., 2016)). As a consequence, when the goal is to infer the species tree, a comparison of orthology inference methods (everything else in the analytical pipeline being unchanged) appears as the most appropriate alternative to assess the robustness of the resulting topology. Yet few studies have compared how sets of orthologs inferred through different methods vary and how it affects species tree reconstruction. Shen *et al.* (2018) compared the performance of OrthoMCL (Li et al., 2003) refined with PhyloTreePruner (Kocot et al., 2013), BUSCO v.2.0.1 (Waterhouse et al., 2017) and PhylomeDB v4 (Huerta-Cepas et al., 2014) in a data set composed of 332 budding yeast (Saccharomycotina) genomes. They compared the overlap between the refined OrthoMCL orthologous groups (with a size of 2,408 orthologous groups, referred to as OGs hereafter) with the BUSCO and PhylomeDB ones, respectively. From a total of 1,292 BUSCO OGs, a large majority were recovered by OrthoMCL as well (1,081 OGs). However, OrthoMCL recovered less than half of the PhylomeDB OGs (819 out of 1,838). Overall, the resulting topologies after the analysis of the concatenated data sets differed in 10% of the nodes (32 out of 331 nodes).

In two studies dealing with spiders interrelationships (a much shallower systematic level than the previous example), Fernández *et al.* (2018) and Kallal *et al.* (2018) compared OGs inferred by BUSCO v1.1b (Simão et al., 2015) and UPhO (Ballesteros and Hormiga, 2016). Contrarily to the example of Shen et al. (2018), these authors did not analyzed the matrix resulting from the intersection of both orthology inference methods, but the BUSCO OGs and UPhO OGs individually. Both studies found congruence between most of the analyses in the concatenated matrices, with minimal topological effects from orthology assessment despite recovering an overlap of as low as 4.3% of OGs between both methods, as reported in Kallal *et al.* (2018).

Finally, Altenhoff *et al.* (2018b) compared several orthology methods (OMA, OrthoMCL, OrthoFinder, HaMStR, and BUSCO) on a reconstruction of the Lophotrochozoa phylogeny. The number of orthologous groups recovered varied quite substantially—ranging from 384 (BUSCO) to 2162 (OMA). Furthermore, the accuracy and branch support of Bayesian and Maximum Likelihood trees reconstructed from these groups varied considerably, suggesting that for difficult phylogenies such as Lophotrochozoa, the choice of orthology inference method can lead to different conclusions.

All in all, while Fernández *et al.* (2018) and Kallal *et al.* (2018) found congruence between most topologies despite differences in the sequences used to infer the species trees—therefore suggesting strong signal in the data robust to differences in orthology inference— Shen *et al. (2018)* and Altenhoff *et al.* (2018b) found that as much as 10% of the nodes varied between topologies. These results highlight the importance of comparing orthology inference methods in each data set as they may strongly affect the resulting species tree topology.

The selection of a proper orthology inference software is of particular importance in complex evolutionary scenarios where gene and genome duplications are frequent, as is the case in plants. Orthology inference methods developed without explicit consideration for such duplication events, such as OrthoMCL (Li et al., 2003), have been reported to be potentially problematic in plants because they tend to break gene families apart instead of retaining its structure (McKain et al., 2018). Instead, other methods better able to account for gene duplications have been recommended in this challenging phylogenomic scenarios, for example OrthoFinder (Emms and Kelly, 2015), OMA (Altenhoff et al., 2018b), PhylomeDB (Huerta-Cepas et al., 2014) or all‑by‑all BLAST followed by Markov clustering and tree‑based orthology pruning (McKain et al., 2018; Yang and Smith, 2014)

Regardless of the software selected for orthology inference, the inclusion of paralogous sequences may result in different outcomes. In some cases, such as in shallow-level phylogenies (e.g., at the level of order, genus, etc.), species tree reconstruction may not be affected by paralogs as far as they are recent enough to be monophyletic for each lineage. In other cases, paralogs have been even proven useful as additional loci for phylogenetics, as reads from the two paralogous sequences can be sorted and assembled into separate, orthologous alignments when the relative age of a genome duplication is known (Johnson et al., 2016).

## Conclusions

As we have reviewed in this chapter, orthology is a fundamental concept for phylogenomics. The terminology, its implications, and the daunting array of methods led to some confusion in the early days of genomics. This has noticeably improved, in large part thanks to a sustained community effort around the *Quest for Orthologs* consortium (Dessimoz et al., 2012; Forslund et al., 2017; Gabaldón et al., 2009; Sonnhammer et al., 2014).

Yet challenges remain. In the context of more than two species, the concept of an orthologous group remains often imprecise in the literature; we have yet to attain the same level of understanding for groupwise orthology as for pairwise orthology. Comparisons among methods has also mainly focused on pairwise orthology. But phylogenomic tree inference requires groups, and several recent studies have observed substantial differences in the trees obtained from different orthologous group reconstruction techniques. Thus, to resolve difficult phylogenies, it may be necessary to better understand and characterise the impact of orthology on tree inference.

## Acknowledgements

CD acknowledges support by Swiss National Science Foundation grant 150654. TG acknowledges support from the European Union's Horizon 2020 research and innovation programme under the grant agreement ERC-2016-724173, and from the Spanish Instituto Nacional de Bioinformática (INB) grant PT17/0009/0023 - ISCIII-SGEFI/ERDF. RF was supported by a Marie Skłodowska-Curie fellowship (grant agreement 747607).